\documentclass[10pt,a4paper]{article}
\usepackage[top=1 in,bottom=1 in,left=1.0 in,right=1.0 in]{geometry}
\usepackage{amsmath,amssymb,amsfonts,amsthm,float,ulem}
\usepackage{cite}
\usepackage{graphicx}
\usepackage{subfigure}
\usepackage{epic,xcolor}
\usepackage[colorlinks,linkcolor=blue,citecolor=blue]{hyperref}
\newtheorem{theorem}{Theorem}
\newtheorem{proposition}{Proposition}
\def \W#1{\widehat{#1}}

\title{Solutions to integrable space-time shifted nonlocal equations}

\author{Shi-min Liu,~ Jing Wang,~ Da-jun Zhang\footnote{
Corresponding author. Email: djzhang@staff.shu.edu.cn} \\
Department of Mathematics, Shanghai University, Shanghai 200444, P.R. China}

\date{\today}

\begin{document}
\maketitle

\begin{abstract}
  In this paper we present a reduction technique based on
  bilinearization and double Wronskians (or double Casoratians) to obtain explicit multi-soliton solutions
  for the integrable space-time shifted nonlocal equations introduced very recently by
  Ablowitz and Musslimani in [Phys. Lett. A, 2021].
  Examples include the space-time shifted nonlocal nonlinear Schr\"odinger and modified Korteweg-de Vries hierarchies
  and the semi-discrete nonlinear Schr\"odinger equation.
  It is shown that these nonlocal integrable equations with or without space-time shift(s) reduction
  share same distributions of eigenvalues but the space-time shift(s) brings new
  constraints to phase terms in solutions.

\begin{description}
%\item[MSC:] 35Q51, 35Q55
%\item[PACS numbers:]
%02.30.Ik, 02.30.Ks, 05.45.Yv
\item[Keywords:]
integrable space-time shifted nonlocal equation, solution,
reduction, bilinear,  double Wronskian
\end{description}
\end{abstract}

%\tableofcontents

\section{Introduction}\label{sec-1}

In a recent paper \cite{AM-PLA-2021} Ablowitz and Mussilimani introduced
integrable space-time shifted nonlocal equations as a generalization of
the nonlocal integrable equations proposed by them in \cite{AM-PRL-2013},
where they introduced nonlocal space reduction $r(x,t)=\pm q^{*}(-x,t)$ and get
a \textit{PT}-symmetric nonlinear Schr\"odinger (NLS) equation as a new and unusual type of integrable equations.
After the pioneer paper \cite{AM-PRL-2013}, the nonlocal reductions have been extended to
inverse time case \cite{AM-Nonl-2016,AM-SAPM-2017}
and many coupled integrable equations, from the continuous to the fully discrete,
were found to allow nonlocal reductions, e.g. \cite{AM-Nonl-2016,AM-SAPM-2017,
AM-PRE-2014,samra-2014,Fokas-Nonl-2016,Gurses-PLA-2017,
SXZ-JPSJ-2017,Zhou-SAPM-2018,CLZ-AML-2019,Lou-SAPM-2019,ZZ-AML-2019,Liu-ROMP-2020,ZVZ-SIGMA-2020}.
In addition, nonlocal integrable systems have  received attention from many aspects,
such as physics backgrounds \cite{LH-SP-2017,Lou-CTP-2020,Yang-PRE-2018,AblM-JPA-2019},
complete integrability \cite{GerS-JMP-2017},
variety of techniques for finding solutions\cite{AM-PRL-2013,AM-Nonl-2016,SXZ-JPSJ-2017,Zhou-SAPM-2018,
AblFLM-SAPM,YY-SAPM-2018,Caud-SAPM-2018,ChenZ-AML-2018,ChenDLZ-SAPM-2018,GP-JMP-2018,
YY-LMP-2019,FZ-ROMP-2019,AblLM-Nonl-2020,RaoCPMH-PD-2020,MS-TMP-2020,ZY-PD-2020},
long-time asymptotics for special initial data \cite{RybS-JDE-2021,RybS-CMP-2021},
and so forth.

As for solutions, nonlocal reductions bring new constraints on eigenvalues of the corresponding spectral problem
and make different eigenvalue distributions from the case of normal reductions.
A recent development of solving nonlocal integrable systems is a direct reduction technique
based on bilinearizations and double Wronskians \cite{ChenZ-AML-2018,ChenDLZ-SAPM-2018}.
This method allows us to make use of known solutions of the unreduced coupled equations and present
explicit solutions to the reduced equations according to canonical forms of the eigenvalue matrix.
In this reduction approach, one can clearly see the differences of the distributions of eigenvalues
in different reductions. Such a method has proved effectively in presenting solutions to
many nonlocal integrable equations, cf.\cite{ChenZ-AML-2018,ChenDLZ-SAPM-2018,DengLZ-AMC-2018,ShiY-ND-2019,
WangWZ-CTP-2020,LWZ-arxiv-2021}.

In this paper we will apply the reduction technique to the integrable space-time shifted nonlocal equations.
Examples we will employ to demonstrate this technique include
the  space-time shifted nonlocal NLS hierarchy, modified Korteweg-de Vries (mKdV) hierarchy,
and the semi-discrete NLS equation.
We will see that, compared with the nonlocal equations without space-time shifts,
the  distributions of eigenvalues do not change but space-time shifts
will bring new constraints to the phase terms in solutions.

The paper is arranged as follows.
In Sec.\ref{sec-2} we present a reduction technique based on bilinearization and double Wronskians
of the unreduced Ablowitz-Kaup-Newell-Segur (AKNS) hierarchy to obtain explicit multi-soliton solutions
for the integrable space-time shifted nonlocal
NLS and mKdV hierarchies, respectively.
In Sec.\ref{sec-3} we start from bilinear form and double Casoratians of the unreduced Ablowitz-Ladik (AL) system
and implement the reduction technique to get solutions for the space-time shifted nonlocal semi-discrete NLS  equation.
Finally, conclusions are given in Sec.\ref{sec-4}.

\section{Solutions to the space-time shifted nonlocal NLS  and mKdV hierarchies}\label{sec-2}

In this section, we will illustrate  nonlocal integrable equations with or without space-time shift(s) reduction
  share same distributions of eigenvalues but the space-time shift(s) brings new
  constraints to phase terms in solutions.
Examples employed are the nonlocal hierarchies related to the AKNS spectral problem
\cite{AKNS-PRL-1973,Abl1974}
\begin{equation}\label{akns-spectral}
 \Phi_x =\left( \begin{array}{cc}
                                  \lambda & q \\
                                  r & -\lambda
                                \end{array}
                              \right)  \Phi,~~\Phi=\left(
                                \begin{array}{c}
                                  \phi_1 \\
                                  \phi_2
                                \end{array}
                              \right),
\end{equation}
where $\lambda$ is the spectral parameter and $u=(q(x,t),r(x,t))^T$ is the vector of two potential functions.

In the following we will first list some shifted nonlocal reductions of the AKNS hierarchy,
and double Wronskian solutions of the unreduced NLS and mKdV hierarchies.
Then we will demonstrate the reduction technique for the space-time shifted nonlocal case.
Finally, we will present explicit solutions for the reduced equations
and as an example illustrate dynamics of solutions for a space-time shifted nonlocal mKdV equation.

\subsection{Space-time shifted nonlocal NLS  and mKdV hierarchies}\label{sec-2-1}

The well known AKNS  hierarchy related to \eqref{akns-spectral} is
\begin{equation}\label{akns-hierarchy}
 u_{t_n}    =K_n     = \left(
                      \begin{array}{c}
                        K_{1,n} \\
                        K_{2,n}
                      \end{array}
                    \right)
                    =L^{n-1} u_x,
                    ~~n=1,2,\cdots,
\end{equation}
with recursion operator
\begin{equation*}
%\label{akns-recusion-operator}
 L= \left(
            \begin{array}{cc}
              -\partial_x+2q \partial_x^{-1}r & 2q \partial_x^{-1}q \\
              -2r \partial_x^{-1}r & \partial_x- 2r \partial_x^{-1}q
            \end{array}
          \right),
\end{equation*}
where $\partial_x=\frac{\partial}{\partial x}$ and  $\partial^{-1}_x$ specially takes the form
\begin{equation}\label{int}
\partial^{-1}_x =\frac{1}{2}(\int^x_{-\infty}-\int^{+\infty}_x)~\cdot~\mathrm{d}x
\end{equation}
so that it is ready for making reverse space reductions.
For the sake of reduction, the hierarchy \eqref{akns-hierarchy} is separated into two:
the even-order hierarchy, i.e., the NLS hierarchy (with $t_{2l}$ replaced by $it_{2l}$
where $i^2=-1$),
\begin{equation}\label{even-hie}
  iu_{t_{2l}}=-K_{2l},~~l=1,2,\cdots,
\end{equation}
and the odd-order hierarchy, i.e., the mKdV hierarchy,
  \begin{equation}\label{odd-hie}
    u_{t_{2l+1}}=K_{2l+1},~ ~l=0,1,2,\cdots.
  \end{equation}
The first system in the NLS hierarchy is
\begin{equation}\label{couple-nls-equ}
  \begin{split}
    & iq_{t_2}= q_{xx}-2q^2 r,\\
    &ir_{t_2}= -r_{xx}+2 qr^2,
  \end{split}
\end{equation}
which allows the following space-time shifted nonlocal reductions \cite{AM-PLA-2021}
\begin{subequations}\label{nls-hie-red}
\begin{align}
  &r(x,t)= \delta q^*(x_0-x,t),~~\delta=\pm 1, \label{nls-hie-reduction1}\\
  &r(x,t)= \delta q(x_0-x,t_0-t),~~\delta=\pm 1, \label{nls-hie-reduction2}\\
  &r(x,t)= \delta q(x,t_0-t),~~\delta=\pm 1, \label{nls-hie-reduction3}
  \end{align}
\end{subequations}
where $*$ stands for complex conjugate.
The first nonlinear  system in the  mKdV hierarchy is
\begin{equation}\label{couple-mkdv-equ}
  \begin{split}
    & q_{t_3}= q_{xxx}-6qq_xr,\\
    &r_{t_3}= r_{xxx}-6qrr_x,
  \end{split}
\end{equation}
which allows space-time shifted nonlocal reductions \cite{AM-PLA-2021}
\begin{subequations}\label{mkdv-hie-redd}
  \begin{align}
    & r(x,t)= \delta q(x_0-x,t_0-t),~~\delta=\pm 1, \label{mkdv-hie-red}\\
    & r(x,t)= \delta q^*(x_0-x,t_0-t),~~\delta=\pm 1.\label{cmkdv-hie-red}
  \end{align}
\end{subequations}

For reductions of the above two hierarchies, we have the following.
\begin{proposition}\label{P-1}
With the definition \eqref{int} for $\partial_x^{-1}$, the reductions \eqref{nls-hie-red}
and \eqref{mkdv-hie-redd} can be extended to the NLS hierarchy \eqref{even-hie} and the mKdV hierarchy \eqref{odd-hie},
respectively.
\end{proposition}

\subsection{Solutions to the unreduced systems}\label{sec-2-2}

The AKNS hierarchy \eqref{akns-hierarchy} can be bilinearized, via
\begin{equation}\label{akns-transformation}
 q=\frac{g}{f},~~r=-\frac{h}{f},
\end{equation}
to \cite{newell1985}
\begin{subequations}\label{akns-bilibear}
\begin{eqnarray}
  &&(D_{t_{n+1}}+D_{x}D_{t_n})g\cdot f =0, \label{akns-bilibear-1}  \\
  &&(D_{t_{n+1}}-D_{x}D_{t_n})h\cdot f =0 , \label{akns-bilibear-2}  \\
  &&D^2_{x}f\cdot f =2gh,\label{akns-bilibear-3}
\end{eqnarray}
\end{subequations}
where $t_1=x$ and $D$ is the Hirota bilinear operator defined by\cite{Hirota-1974}
\begin{equation*}
 D_x^m D_y^n f(x,y)\cdot g(x,y) = (\partial_x- \partial_{x'})^m(\partial_y- \partial_{y'})^n f(x,y)g(x',y')|_{x'=x,y'=y}.
\end{equation*}
The above hierarchy admits double Wronskian solutions \cite{liu1990,yin2008}.

Recalling the results given in \cite{ChenDLZ-SAPM-2018,yin2008}, we directly list solutions to the unreduced hierarchies
\eqref{even-hie} and \eqref{odd-hie}.

\begin{proposition}\label{P-2}
(1). Solutions to the NLS hierarchy \eqref{even-hie} are given by \eqref{akns-transformation} where
\begin{equation}\label{anks-solution-fgh}
f= |\widehat{\varphi^{[N]}}; \widehat{\psi^{[M]}}|,~~g= 2|\widehat{\varphi^{[N-1]}_{}}; \widehat{\psi^{[M+1]}}|,
~~h= 2|\widehat{\varphi^{[N+1]}}; \widehat{\psi^{[M-1]}}|,
\end{equation}
where  $\varphi$ and $\psi$ are respectively  $(N+M+2)$-th order column vectors
\begin{equation}\label{akns-var-psi-A-c-d-even}
\varphi=\exp{\Bigl(\frac{1}{2}Ax+\frac{i}{2}\sum^{\infty}_{j=1} A^{2j} t_{2j}\Bigr)}C^+, ~~
\psi=\exp{\Bigl( -\frac{1}{2}Ax-\frac{i}{2}\sum^{\infty}_{j=1} A^{2j} t_{2j}\Bigr)}C^-,
\end{equation}
with arbitrary $A\in \mathbb{C}_{(N+M+2)\times (N+M+2)}$ and $C^{\pm}\in \mathbb{C}_{(N+M+2)}$,
and the shorthand  $\widehat{\varphi^{[N]}}$ denotes Wronski matrix
\begin{equation}\label{tag-hat-varphi}
\widehat{\varphi^{[N]}}=\widehat{\varphi^{[N]}}(x)=(\varphi(x), \partial_x\varphi(x),
\partial^2_x\varphi(x),\cdots,\partial^{N}_x\varphi(x)).
\end{equation}
(2). Solutions to the mKdV hierarchy \eqref{odd-hie} are given by \eqref{akns-transformation} with double Wronskians
\eqref{anks-solution-fgh}, composed by
\begin{equation}\label{akns-var-psi-A-c-d-odd}
\varphi=\exp{\Bigl(\frac{1}{2}Ax+\frac{1}{2}\sum^{\infty}_{j=0} A^{2j+1} t_{2j+1}\Bigr)}C^+, ~~
\psi=\exp{\Bigl( -\frac{1}{2}Ax-\frac{1}{2}\sum^{\infty}_{j=0} A^{2j+1} t_{2j+1}\Bigr)}C^-.
\end{equation}
(3). Matrix $A$ and any of its similar form lead to same $q$ and $r$.
\end{proposition}

\subsection{Reductions of solutions to the space-time shifted  nonlocal  cases}\label{sec-2-3}

Let us take the reduction \eqref{nls-hie-reduction1}, i.e. $r(x,t)= \delta q^*(x_0-x,t)$, as an example,
to show how the reduction technique works.

Let $M=N$  in \eqref{anks-solution-fgh} and impose constraint
\begin{subequations}\label{nls1-psiTphic}
  \begin{align}
 &\psi(x,t)=T \varphi^{*}(x_0-x,t),\label{nls1-psiTphi}\\
 &C^-=Te^{\frac{1}{2}A^*x_0}C^{+*},\label{nls1-ct}
  \end{align}
\end{subequations}
where $T\in \mathbb{C}_{2(N+1)\times 2(N+1)}$ is a constant matrix determined by the system
\begin{subequations}\label{nls1-at-t-equation}
  \begin{align}
&AT- TA^*=0 ,\label{nls1-at-equation}\\
 &TT^*=-\delta I,\label{nls1-tt-equation}
\end{align}
\end{subequations}
where $I$ is the $2(N+1)$-th order unit matrix.
In fact, when   \eqref{nls1-ct} and \eqref{nls1-at-equation} hold,
we have
\begin{align*}
\psi(x,t)&= \exp{\Bigl( -\frac{1}{2}Ax-\frac{i}{2}\sum^{\infty}_{j=1} A^{2j} t_{2j}\Bigr)}C^-    \\
&=  \exp{\Bigl( -\frac{1}{2}(T A^*T^{-1})x-\frac{i}{2}\sum^{\infty}_{j=1} ( T A^*T^{-1})^{2j} t_{2j}\Bigr)}
T\exp{\Bigl(\frac{1}{2}A^*x_0\Bigr)}C^{+*}    \\
&=  T\exp{\Bigl( -\frac{1}{2} A^*x -\frac{i}{2}\sum^{\infty}_{j=1} { A^{*}}^{2j}t_{2j}\Bigr)}
\exp{\Bigl(\frac{1}{2}A^*x_0\Bigr)}C^{+*}     \\
&=  T\exp{\Bigl( \frac{1}{2} A^*(x_0-x) - \frac{i}{2} \sum^{\infty}_{j=1}  {A^{*}}^{2j }t_{2j}\Bigr )}C^{+*}    \\
&= T \varphi^*(x_0-x,t),
\end{align*}
which indicates \eqref{nls1-psiTphi} is valid.
Next, in order to examine relations between double Wronskians, based on \eqref{tag-hat-varphi}, let us introduce a notation
\begin{align*}
  &\widehat{\varphi^{[N]}}(a(x))_{[b(x)]} = \Bigl(\varphi(a(x)),\partial_{b(x)}\varphi(a(x)),\partial_{b(x)}^2\varphi(a(x)),
  \cdots,\partial_{b(x)}^{N}\varphi(a(x))\Bigr)
\end{align*}
where $a(x)$ and $b(x)$ are functions of $x$.
Then, with the constraint \eqref{nls1-psiTphi} and $N=M$, the double Wronskians $f,g,h$ in \eqref{anks-solution-fgh}
are rewritten as
\begin{subequations}\label{nls1-fgh}
\begin{eqnarray}
&&f(x,t)= |\widehat{\varphi^{[N]}}; \widehat{\psi^{[N]}}|= |\widehat{\varphi^{[N]}}(x,t)_{[x]};
T\widehat{\varphi^{*{[N]}}}(x_0-x,t)_{[x]}|,  \label{nls1-fgh-f}  \\
&&g(x,t)= 2|\widehat{\varphi^{[N-1]}}; \widehat{\psi^{[N+1]}}|=  2|\widehat{\varphi^{[N-1]}}(x,t)_{[x]};
T\widehat{\varphi^{*{[N+1]}}}(x_0-x,t)_{[x]}|,  \label{nls1-fgh-g}  \\
&&h(x,t)= 2|\widehat{\varphi^{[N+1]}}; \widehat{\psi^{[N-1]}}|=  2|\widehat{\varphi^{[N+1]}}(x,t)_{[x]};
T\widehat{\varphi^{*{[N-1]}}}(x_0-x,t)_{[x]}|. \label{nls1-fgh-h}
\end{eqnarray}
\end{subequations}
With a further assumption \eqref{nls1-tt-equation},
we find
\begin{align*}
f^*(x_0-x,t)&= |\widehat{\varphi^{*{[N]}}}(x_0- x,t)_{[x_0-x]}; T^*\widehat{\varphi^{[N]}}(x,t)_{[x_0- x]}|    \\
& = |T^*| |T^{*^{-1}}\widehat{\varphi^{*{[N]}}}(x_0- x,t)_{[x]}; \widehat{\varphi^{[N]}}(x,t)_{[x]}|   \\
& = (-1)^{(N+1)^2}|T^*| | \widehat{\varphi^{[N]}}(x,t)_{[x]};T^{*^{-1}}\widehat{\varphi^{*{[N]}}}(x_0- x,t)_{[x]}|  \\
& = (-1)^{(N+1)^2}|T^*| | \widehat{\varphi^{[N]}}(x,t)_{[x]};-\delta T\widehat{\varphi^{*{[N]}}}(x_0- x,t)_{[x]}|  \\
&= (-1)^{(N+1)^2+N+1}\delta^{N+1}|T^*| | \widehat{\varphi^{[N]}}(x,t)_{[x]};
T\widehat{\varphi^{*{[N]}}}(x_0- x,t)_{[x]}| \\
&= \delta^{N+1}|T^*| f(x,t),
\end{align*}
and in the same way, $g^*(x_0-x,t)=-\delta^{N}|T^*| h(x,t)$,
which then indicates
\begin{equation*}
\delta q^*(x_0-x,t)= \delta\frac{ g^*(x_0-x,t)}{f^*(x_0-x,t)}
= \frac{-\delta^{N+1}|T^*| h(x,t)}{\delta^{N+1}|T^*| f(x,t)}=-\frac{h(x,t)}{f(x,t)}=r(x,t),
\end{equation*}
i.e. the reduction \eqref{nls-hie-reduction1} holds.

In a similar way, we can check the reductions \eqref{nls-hie-reduction2} and
\eqref{nls-hie-reduction3} by imposing
\[C^-=T\exp{\Bigl(\frac{1}{2}Ax_0+\frac{i}{2}\sum^{\infty}_{j=1}A^{2j}t_0\Bigr)}C^{+}\]
and
\[C^-=T\exp{\Bigl(\frac{i}{2}\sum^{\infty}_{j=1}A^{2j}t_0\Bigr)}C^{+},\]
respectively.

In the following we skip details of  proof and present the reduced solutions for the
space-time  shifted nonlocal  NLS hierarchy
\begin{equation}\label{nls1-hie}
  iq_{t_{2l}}=- K_{1,2l} |_{(\ref{nls-hie-red}*)}, ~~l=1,2,\cdots,
 \end{equation}
where (\ref{nls-hie-red}$*$) means either \eqref{nls-hie-reduction1} or \eqref{nls-hie-reduction2} or
\eqref{nls-hie-reduction3}.

\begin{theorem}\label{T-1}
The reduced NLS hiererchy \eqref{nls1-hie} admit solutions in the form
\begin{equation}\label{nls1-q-solu}
 q(x,t)= \frac{2|\widehat{\varphi^{[N-1]}}; \widehat{\psi^{[N+1]}}|}{|\widehat{\varphi^{[N]}}; 
 \widehat{\psi^{[N]}}|},
\end{equation}
where the $2(N+1)$-th order column vector $\varphi$ takes the form
\begin{equation}\label{rewrite-akns-var-psi-A-c-d-even-1}
  \varphi=\exp{\Bigl(\frac{1}{2}Ax+\frac{i}{2}\sum^{\infty}_{j=1} A^{2j} t_{2j}\Bigr)}C^+.
\end{equation}
For the reduction \eqref{nls-hie-reduction1}, i.e. $r(x,t)= \delta q^*(x_0-x,t)$,
\begin{equation}\label{nls1-psiT-1}
\psi(x,t)=T \varphi^{*}(x_0-x,t)
\end{equation}
in which $T$ is governed by the system
\begin{equation}\label{nls1-at-t-eq}
AT- TA^*=0 ,~~TT^*=-\delta I.
\end{equation}
For the reduction \eqref{nls-hie-reduction2}, i.e. $r(x,t)= \delta q(x_0-x,t_0-t)$,
\begin{equation}
\psi(x,t)=T \varphi(x_0-x, t_0- t),\label{nls2-psiTphi}
\end{equation}
in which $T$ is governed by
\begin{equation}\label{nls2-at-t-equation}
AT- TA=0,~~ T^2=-\delta I.
\end{equation}
For  the reduction \eqref{nls-hie-reduction3}, i.e. $r(x,t)= \delta q(x,t_0-t)$,
\begin{equation}
\psi(x,t)=T \varphi(x,t_0-t)
\end{equation}
in which $T$ is governed by
\begin{equation}\label{nls3-at-t-equation}
 AT+TA=0, ~~ T^2=\delta I.
\end{equation}
\end{theorem}

In a similar way, we can also obtain the reduced solutions for the
space-time  shifted nonlocal  mKdV hierarchy
\begin{equation}\label{mkdv-hie}
  q_{t_{2l+1}}= K_{1,2l+1} |_{(\ref{mkdv-hie-redd}*)}, ~~l=0,1,2,\cdots,
 \end{equation}
where (\ref{mkdv-hie-redd}$*$) means either \eqref{mkdv-hie-red} or \eqref{cmkdv-hie-red}.

\begin{theorem}\label{T-2}
The reduced mKdV hiererchy \eqref{mkdv-hie} admit solutions
\begin{equation}\label{mkdv-q-solu}
 q(x,t)= \frac{2|\widehat{\varphi^{[N-1]}}; \widehat{\psi^{[N+1]}}|}
 {|\widehat{\varphi^{[N]}}; \widehat{\psi^{[N]}}|},
\end{equation}
where the $2(N+1)$-th order column vector $\varphi$ takes the form
\begin{equation}\label{rewrite-akns-var-psi-A-c-d-odd-1}
  \varphi=\exp{\Bigl(\frac{1}{2}Ax+\frac{1}{2}\sum^{\infty}_{j=0} A^{2j+1} t_{2j+1}\Bigr)}C^+.
\end{equation}
For the reduction \eqref{mkdv-hie-red}, i.e. $r(x,t)= \delta q(x_0-x,t_0-t)$,
\begin{equation}
\psi(x,t)=T \varphi(x_0-x,t_0-t),\label{mkdv-psiTphi}
\end{equation}
in which $T$ is governed by the system \eqref{nls2-at-t-equation}. 
For the reduction \eqref{cmkdv-hie-red}, i.e. $r(x,t)= \delta q^*(x_0-x,t_0-t)$,
\begin{equation}
\psi(x,t)=T \varphi^*(x_0-x,t_0-t),\label{cmkdv-psiTphi}
\end{equation}
in which $T$ is governed by the system \eqref{nls1-at-t-eq}. 
\end{theorem}

Note that for reduction \eqref{mkdv-hie-red} and \eqref{cmkdv-hie-red}, $C^-$ takes as
\[C^-=T\exp{\Bigl(\frac{1}{2}Ax_0+\frac{1}{2}\sum^{\infty}_{j=0}A^{2j+1}t_0\Bigr)}C^{+}\]
and
\[C^-=T\exp{\Bigl(\frac{1}{2}A^*x_0+\frac{1}{2}\sum^{\infty}_{j=0}A^{*^{2j+1}}t_0\Bigr)}C^{+*},\]
respectively.

\subsection{Explicit solutions }\label{sec-2-4}

To get explicit solutions one only needs to solve the matrix system
\eqref{nls1-at-t-eq}, \eqref{nls2-at-t-equation} and \eqref{nls3-at-t-equation}
and get explicit forms for matrices $A$ and $T$.
Special solutions to these matrix equations have been obtained before \cite{ChenDLZ-SAPM-2018}
when $A$ and $T$ are block matrices
\begin{align}
 A=\left(\begin{array}{cc}
                   K_1 & \mathbf{0} \\
                   \mathbf{0} & K_4
                 \end{array}
               \right),~~
               T=\left(
     \begin{array}{cc}
       T_1 & T_2 \\
       T_3 & T_4
     \end{array}
   \right),
\label{nls1-example-tac}
\end{align}
where $T_i$ and $K_i$ are $(N+1)\times (N+1)$ matrices.
For convenience we introduce notations
\begin{align}
& \mathrm{Diag}[k_i]_{i=1}^{N+1}\doteq\mathrm{Diag}(k_1, k_2, \cdots, k_{N+1}),\label{K-diag}\\
& \mathbf{J}_{N+1}[k]\doteq \left(
  \begin{array}{cccc}
    k & 0 & \cdots & 0 \\
    1& k & \cdots & 0 \\
    \vdots & \ddots & \ddots & \vdots \\
    0 & \cdots& 1 & k \\
  \end{array}
\right)_{(N+1)\times (N+1)},\label{K-jor}\\
&
  \eta(k )=\left\{\begin{array}{ll}
  \frac{1}{2}kx+\frac{i}{2}\sum^{\infty}_{j=1}k^{2j} t_{2j},& ~(\mathrm{for~ NLS}),\\
   \frac{1}{2}kx+\frac{1}{2}\sum^{\infty}_{j=0}k^{2j+1} t_{2j+1},& ~(\mathrm{for~ mKdV}).
   \end{array}\right.
  \label{eta-nls}
\end{align}

\subsubsection{Solutions corresponding to \eqref{nls1-at-t-eq}}\label{sec-2-4-1}

\begin{proposition}\label{P-3}\cite{ChenDLZ-SAPM-2018}
The system \eqref{nls1-at-t-eq} allows complex solution \eqref{nls1-example-tac} for $\delta=\pm 1$,
where
\begin{equation}\label{case-1-T-A}
  T_{1}=T_{4}=\mathbf{0}_{N+1}, ~T_{2}=-\delta T_{3}=\mathbf{I}_{N+1},~
  K_{1}=K_{4}^{*}=\mathbf{K}_{N+1}\in \mathbb{C}_{(N+1)\times (N+1)},
  \end{equation}
and real solution for only $\delta=-1$, where
\begin{equation}\label{case-2-T-A}
  \begin{split}
  &T_{1}=-T_{4}= \mathbf{I}_{N+1},~~ ~T_{2}=T_{3}=\mathbf{0}_{N+1},\\
  &K_{1}=\mathbf{K}_{N+1} \in \mathbb{R}_{(N+1)\times (N+1)},~~
  K_4=\mathbf{H}_{N+1}  \in \mathbb{R}_{(N+1)\times (N+1)}.
  \end{split}
\end{equation}
Here $\mathbf{I}_{N+1}$ is the unit matrix of order $N+1$.
\end{proposition}

Then, corresponding to \eqref{case-1-T-A}, when
\begin{equation}
\mathbf{K}_{N+1}=\mathrm{Diag}[k_i]_{i=1}^{N+1},~~ k_i\in \mathbb{C},
\end{equation}
an explicit form of  $\varphi$ is
\begin{align}
\varphi=\Bigl(c_{1}e^{\eta(k_{1})},c_{2}e^{\eta(k_{2})}, \cdots, c_{N+1}e^{\eta(k_{N+1})},
d_{1}e^{\eta(k_{1}^*)},d_{2}e^{\eta(k_{2}^*)},\cdots, d_{N+1}e^{\eta(k_{N+1}^*)}\Bigr)^{T},
\end{align}
where $k_i, c_i, d_i\in \mathbb{C}$;
when
\begin{equation}
\mathbf{K}_{N+1}=\mathbf{J}_{N+1}[k],
\end{equation}
$\varphi$ takes
\begin{align}
\varphi=\Bigl(ce^{\eta(k)},\frac{\partial_{k}}{1!}(ce^{\eta(k)}), \cdots,
\frac{\partial_{k}^{N}}{N!}(ce^{\eta(k)}),
de^{\eta (k^*)},\frac{\partial_{k^*}}{1!}(de^{\eta (k^*)}),\cdots,
\frac{\partial_{k^*}^{N}}{N!}(de^{\eta (k^*)})\Bigr)^{T},
\end{align}
where $k,c,d \in \mathbb{C}$.

In the second case \eqref{case-2-T-A}, $A$ is composed by two different real matrices,
$\mathbf{K}_{N+1}$ and $\mathbf{H}_{N+1}$.
In this case, one can take both $\mathbf{K}_{N+1}$ and $\mathbf{H}_{N+1}$ to be diagonals, or Jordan blocks,
or one diagonal and one Jordan block.
For example, when
\begin{equation}\label{KH-diag}
  \mathbf{K}_{N+1}=\mathrm{Diag}[k_i]_{i=1}^{N+1},~ ~
  \mathbf{H}_{N+1}=\mathrm{Diag}[h_i]_{i=1}^{N+1},
  \end{equation}
the corresponding $\varphi$ is
\begin{align}\label{phi-dia}
  \varphi =\Bigl(c_{1}e^{\eta(k_{1})},c_{2}e^{\eta(k_{2})},\cdots,c_{N+1}e^{\eta(k_{N+1})},
  d_{1}e^{\eta(h_1)},d_{2}e^{\eta(h_2)},\cdots,d_{N+1}e^{\eta(h_{N+1})}\Bigr)^{T},
\end{align}
where $c_i,d_i \in \mathbb{C},~k_i,h_i\in \mathbb{R}$;
when
 \begin{equation}\label{KH-D-jordan}
  \mathbf{K}_{N+1}=\mathrm{Diag}[k_i]_{i=1}^{N+1},~ ~\mathbf{H}_{N+1}=\mathbf{J}_{N+1}[h],
 \end{equation}
we have
  \begin{align}\label{phi-D-jor}
  \varphi=\Bigl(c_{1}e^{\eta(k_{1})},c_{2}e^{\eta(k_{2})},\cdots,c_{N+1}e^{\eta(k_{N+1})},
  de^{\eta (h)},\frac{\partial_{h}}{1!}(de^{\eta (h)}),\cdots,
  \frac{\partial_{h}^{N}}{N!}(de^{\eta (h)})\Bigr)^{T},
  \end{align}
where $c_j,d\in \mathbb{C},~k_j,h\in \mathbb{R}$,
and when
 \begin{equation}\label{KH-jordan}
  \mathbf{K}_{N+1}=\mathbf{J}_{N+1}[k], ~\mathbf{H}_{N+1}=\mathbf{J}_{N+1}[h],
 \end{equation}
we have
  \begin{align}\label{phi-jor}
  \varphi=\Bigl(ce^{\eta(k)},\frac{\partial_{k}}{1!}(ce^{\eta(k)}), \cdots,
  \frac{\partial_{k}^{N}}{N!}(ce^{\eta(k)}),
  de^{\eta (h)},\frac{\partial_{h}}{1!}(de^{\eta (h)}),\cdots,
  \frac{\partial_{h}^{N}}{N!}(de^{\eta (h)})\Bigr)^{T},
  \end{align}
where $c,d\in \mathbb{C},~k,h\in \mathbb{R}$.

Note that since \eqref{nls1-at-equation} is a linear equation w.r.t $A$ or $T$, in principle,
one can combine the above cases to get variety of mixed solutions.

\begin{proposition}\label{P-4}
When $\delta=-1$, equation \eqref{nls1-at-t-eq} admits mixed solution
\begin{equation}\label{TA-mix}
 T=\left(
     \begin{array}{cccc}
      \mathbf{I}_{N_1} & \mathbf{0}_{N_1} & & \\
      \mathbf{0}_{N_1}   & -\mathbf{I}_{N_1} & & \\
      & & \mathbf{0}_{N_2}   & \mathbf{I}_{N_2} \\
      & & \mathbf{I}_{N_2} & \mathbf{0}_{N_2}
     \end{array}
   \right),~~
 A=\left(
     \begin{array}{cccc}
      \mathbf{K}'_{N_1} & \mathbf{0}_{N_1} & & \\
      \mathbf{0}_{N_1}   & \mathbf{H}'_{N_1} & & \\
      & & \mathbf{K}_{N_2}   & \mathbf{0}_{N_2} \\
      & & \mathbf{0}_{N_2} & \mathbf{K}^*_{N_2}
     \end{array}
   \right),
\end{equation}
where $\mathbf{K}'_{N_1}, \mathbf{H}'_{N_1} \in \mathbb{R}_{N_1\times N_1}$,
 $\mathbf{K}_{N_2}  \in \mathbb{C}_{N_2\times N_2},~N_1+N_2=N+1$.
\end{proposition}

Obviously,  explicit expression for $\varphi$ of this case can be given accordingly.

\subsubsection{Solutions corresponding to \eqref{nls2-at-t-equation} and \eqref{nls3-at-t-equation}}\label{sec-2-4-2}

Equations \eqref{nls2-at-t-equation} and \eqref{nls3-at-t-equation} can be unified to be
\begin{equation}\label{theorem-non-AT}
 AT+\sigma TA=0,~~T^{2}=\sigma\delta I,~~\sigma,\delta=\pm 1,
  \end{equation}
and then their solutions can be listed out as in the following table.
\begin{table}[h]
  \begin{center}
  \caption{$T$ and $A$ for \eqref{theorem-non-AT} }
  \begin{tabular}{l|c|c}
    \hline
    $(\sigma,\delta)$ & $T$ ~~~~~~~~~~& $A$ \\
    \hline
    $(-1,-1)$ & $T_{1}=- T_{4}=\mathbf{I}_{N+1},T_{3}=T_{2}=\mathbf{0}_{N+1}$
    & $K_{1}=\mathbf{K}_{N+1}, K_{4}=\mathbf{H}_{N+1}$ \\
     \hline
    $(-1,1)$ &$ T_{1}=- T_{4}=i\mathbf{I}_{N+1},T_{3}=T_{2}=\mathbf{0}_{N+1}$
    & $K_{1}=\mathbf{K}_{N+1},K_{4}=\mathbf{H}_{N+1}$ \\
    \hline
    $(1,-1)$ & $T_{1}=T_{4}=\mathbf{0}_{N+1},T_{3}=-T_{2}=\mathbf{I}_{N+1}$
    & $K_{1}=-K_{4}=\mathbf{K}_{N+1}$\label{case-1}\\
    \hline
    $(1,1)$  &$T_{1}=T_{4}=\mathbf{0}_{N+1},T_{3}=T_{2}=\mathbf{I}_{N+1}$
    &$ K_{1}=-K_{4}=\mathbf{K}_{N+1}$ \\
    \hline
  \end{tabular}
    \label{Tab-1}
  \end{center}
  \end{table}

Note that here $K_1, K_4$ are complex. As we have listed in Sec.\ref{sec-2-4-1},
it is easy to write out explicit forms of $\varphi$ of this case.
For example, corresponding to \eqref{nls2-at-t-equation}, i.e. $(\sigma,\delta)=(-1,\pm 1)$,
when  $\mathbf{K}_{N+1}$ and $\mathbf{H}_{N+1}$ take the form \eqref{KH-diag}, \eqref{KH-D-jordan}
and \eqref{KH-jordan}, respectively,  the corresponding $\varphi$ takes the
same form as \eqref{phi-dia}, \eqref{phi-D-jor} and \eqref{phi-jor}, respectively,
but here, the eigenvalues $k_i, h_i, k, h$ are complex.

\subsubsection{An example}\label{sec-2-4-3}

As an example we consider solutions of the complex space-time shifted mKdV equation
$q_{t_3}=K_{1,3} |_{\eqref{cmkdv-hie-red}}$ with $\delta=-1$, i.e.
\begin{equation}
  q_{t_3}=q_{xxx}+6 q q_x q^*(x_0-x,t_0-t_3).
  \label{re-cmkdv-eq}
\end{equation}
It has two one-soliton solutions, which are
\begin{equation}
 q =\frac{c_{1}^{*}d_{1}^{*}(k_1-k_{1}^{*})}{|c_1|^2
 e^{-\frac{1}{2}k_1x_0-\frac{1}{2}k_{1}^3t_0+k_1x
   +k_{1}^{3}t_3}-|d_{1}|^2e^{-\frac{1}{2}k_{1}^*x_0-\frac{1}{2}k_{1}^{*^3}t_0+k_{1}^*x+k_{1}^{*^3}t_3}}
\label{q1-cmkdv-case-1}
\end{equation}
where $k_1,c_1,d_1 \in \mathbb{C}$,
and
\begin{equation}
q=\frac{c_{1}^{*}d_{1}^{*}(k_1-h_1)}{c_1d_{1}^*e^{-\frac{1}{2}k_1x_0
-\frac{1}{2}k_{1}^3t_0+k_1x
   +k_{1}^{3}t_3}+c_{1}^*d_1e^{-\frac{1}{2}h_{1}x_0-\frac{1}{2}h_{1}^3t_0+h_1x+h_{1}^3t_3}}
\label{q1-cmkdv-case-2}
\end{equation}
where $k_1,h_1  \in \mathbb{R}$, $c_1,d_1 \in \mathbb{C}$.
Consider carrier waves of them, we have, respectively,
\begin{equation}\label{abs-q-case-1}
|q|^{2} =\frac{4|c_{1}|^2|d_{1}|^2k_{12}^2e^{k_{11}x_0+(k_{11}^3
-3k_{11}k_{12}^2)t_0-2k_{11}x-2(k_{11}^3-3k_{11}k_{12}^2)t_3}}{|c_{1}|^4
+|d_{1}|^4-2|c_{1}|^2|d_{1}|^2\cos\Bigl(2k_{12}x+2(3k_{11}^2k_{12}-k_{12}^3)t_3
-k_{12}x_0-(3k_{11}^2k_{12}-k_{12}^3)t_0\Bigr)},
\end{equation}
where we have taken $k_1=k_{11}+ik_{12}$ and $k_{11},k_{12}\in \mathbb{R}$, and
\begin{equation}\label{abs-q-case-2}
  |q|^{2} =\frac{(k_{1}-h_1)^2e^{\frac{1}{2}(k_1+h_1)x_0
  +\frac{1}{2}(k_{1}^3+h_{1}^3)t_0-(k_1+h_1)x-(k_{1}^3+h_{1}^3)t_3}}
  {4\cosh^2\Bigl(\frac{1}{2}(k_1-h_1)x+\frac{1}{2}(k_{1}^3-h_{1}^3)t_3-\frac{1}{4}(k_1-h_1)x_0
  -\frac{1}{4}(k_{1}^3-h_{1}^3)t_0\Bigr)},
  \end{equation}
where we have taken $c_1, d_1\in \mathbb{R}$ for convenience.

\begin{figure}[ht]
  \centering
  \subfigure[]{
  \begin{minipage}{6.0cm}%\label{fig1-1a}
  \includegraphics[width=\textwidth]{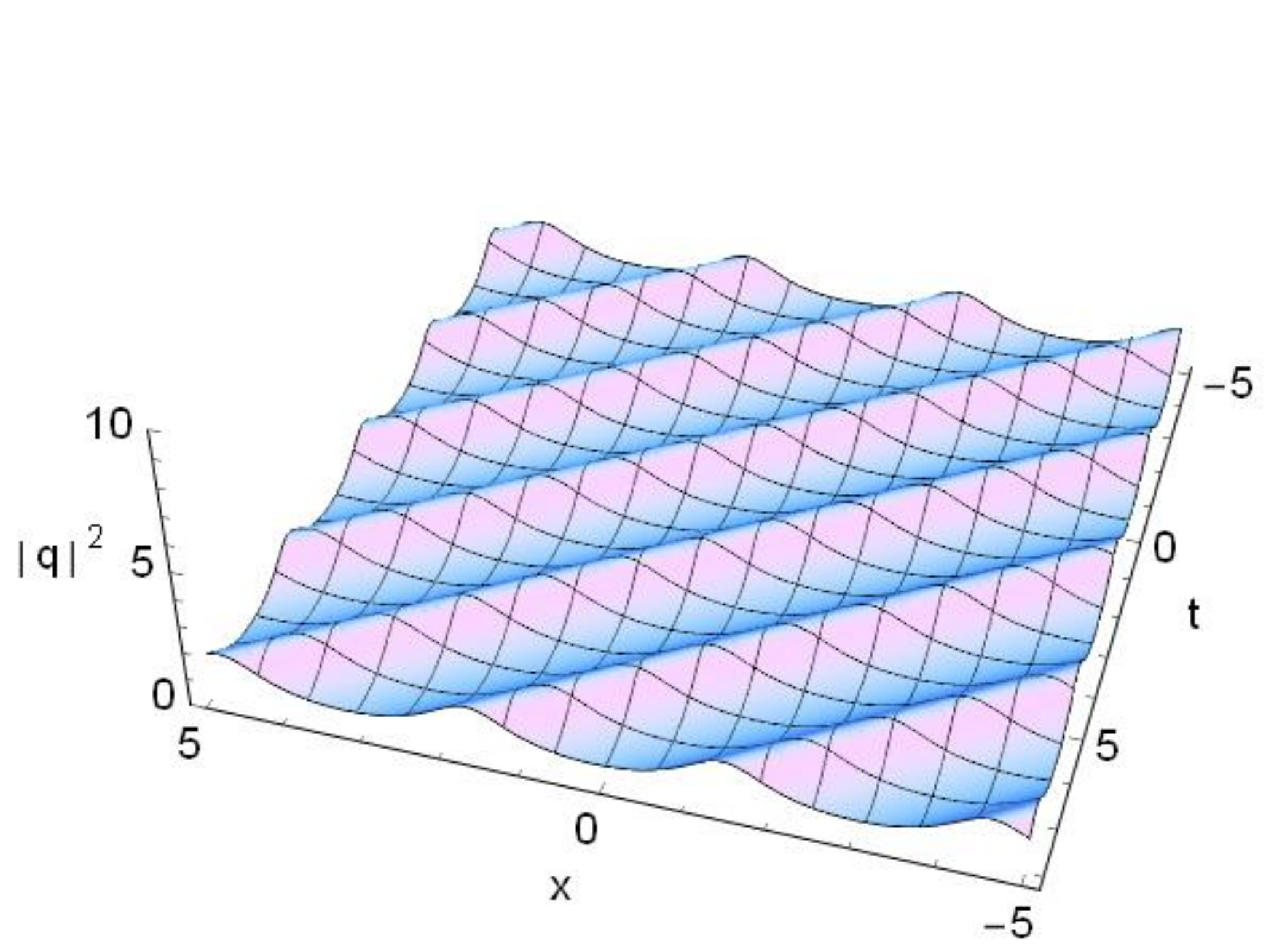}
  \end{minipage}
  }
  \subfigure[]{
  \begin{minipage}{6.0cm}%\label{fig1-1b}
  \includegraphics[width=\textwidth]{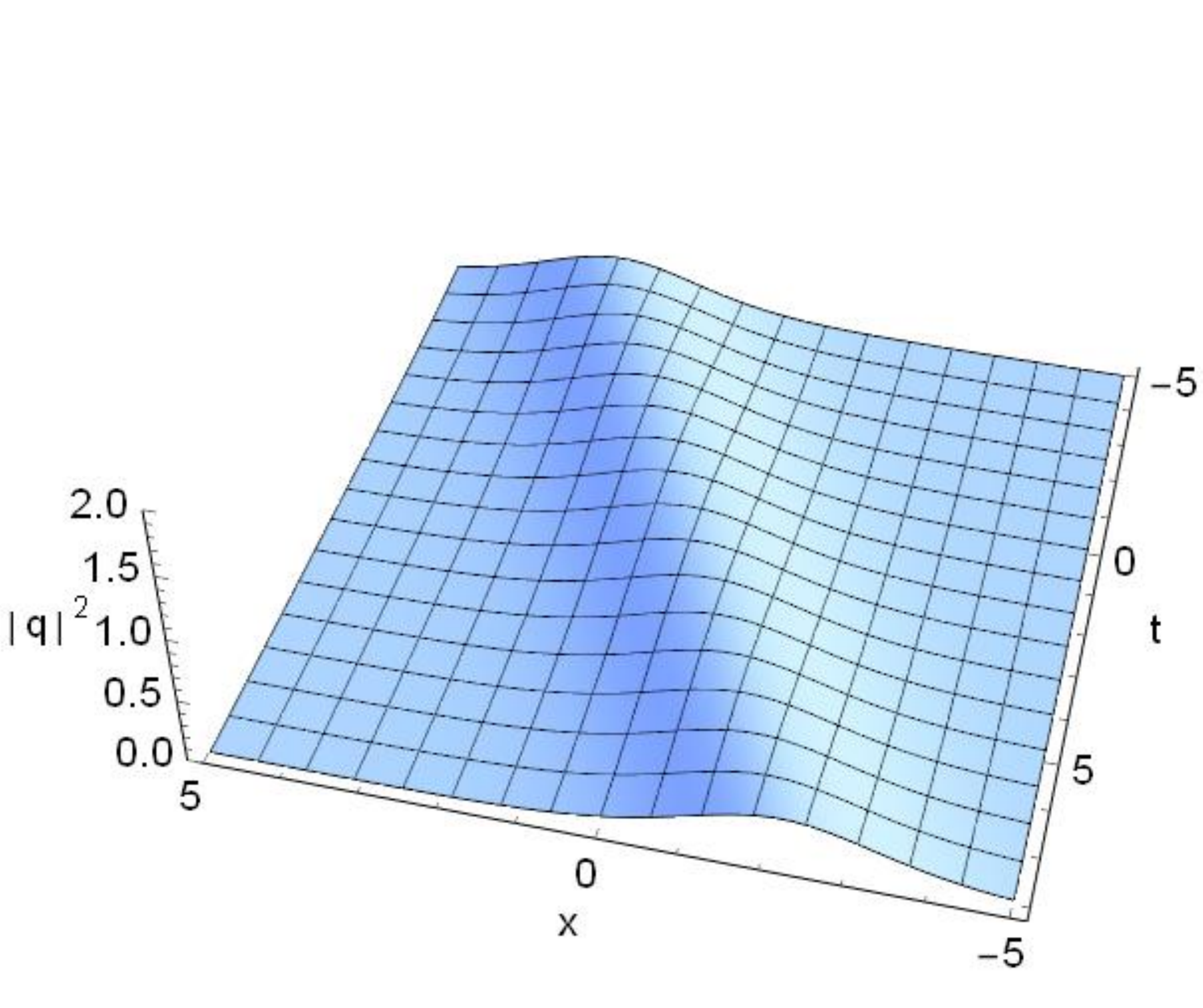}
  \end{minipage}
  }
  \caption{\label{Fig-1} (a). Shape and motion of $|q|^2$ \eqref{abs-q-case-1} for equation \eqref{re-cmkdv-eq},
  in which $k_{11}=0,k_{12}=1, x_0=1,t_0=1, c_1=2,d_1=1$.
     (b). Shape and motion of $|q|^2$ \eqref{abs-q-case-2} for equation \eqref{re-cmkdv-eq},
     in which $k_{1}=0.6,h_{1}=-0.6, x_0=1,t_0=1, c_1=1, d_1=1$.
  }
\end{figure}

For \eqref{abs-q-case-1}, when $k_{11}=0$ but $k_{12}\neq 0$, it provides a periodic wave
as depicted in Fig.\ref{Fig-1}(a). One can find 
that  the `period' (here we mean the distance between two parallel waves) is
\[d=\frac{\pi}{|k_{12}|\sqrt{1+k_{12}^4} }.\]
For \eqref{abs-q-case-2}, it is a moving wave with a $\{x,t_3\}$-dependent amplitude.
However, when taking $k_1=-h_1$, the wave becomes a standard soliton moving with a constant amplitude $k_1^2$
along the vertex trajectory
\[x=-k_{1}^2t_3+\frac{1}{2}x_0+\frac{1}{2}k_{1}^2t_0.\]
We depicted such  a wave in Fig.\ref{Fig-1}(b).

\begin{figure}[ht]
  \centering
  \subfigure[]{
  \begin{minipage}{6.5cm}
  \includegraphics[width=\textwidth]{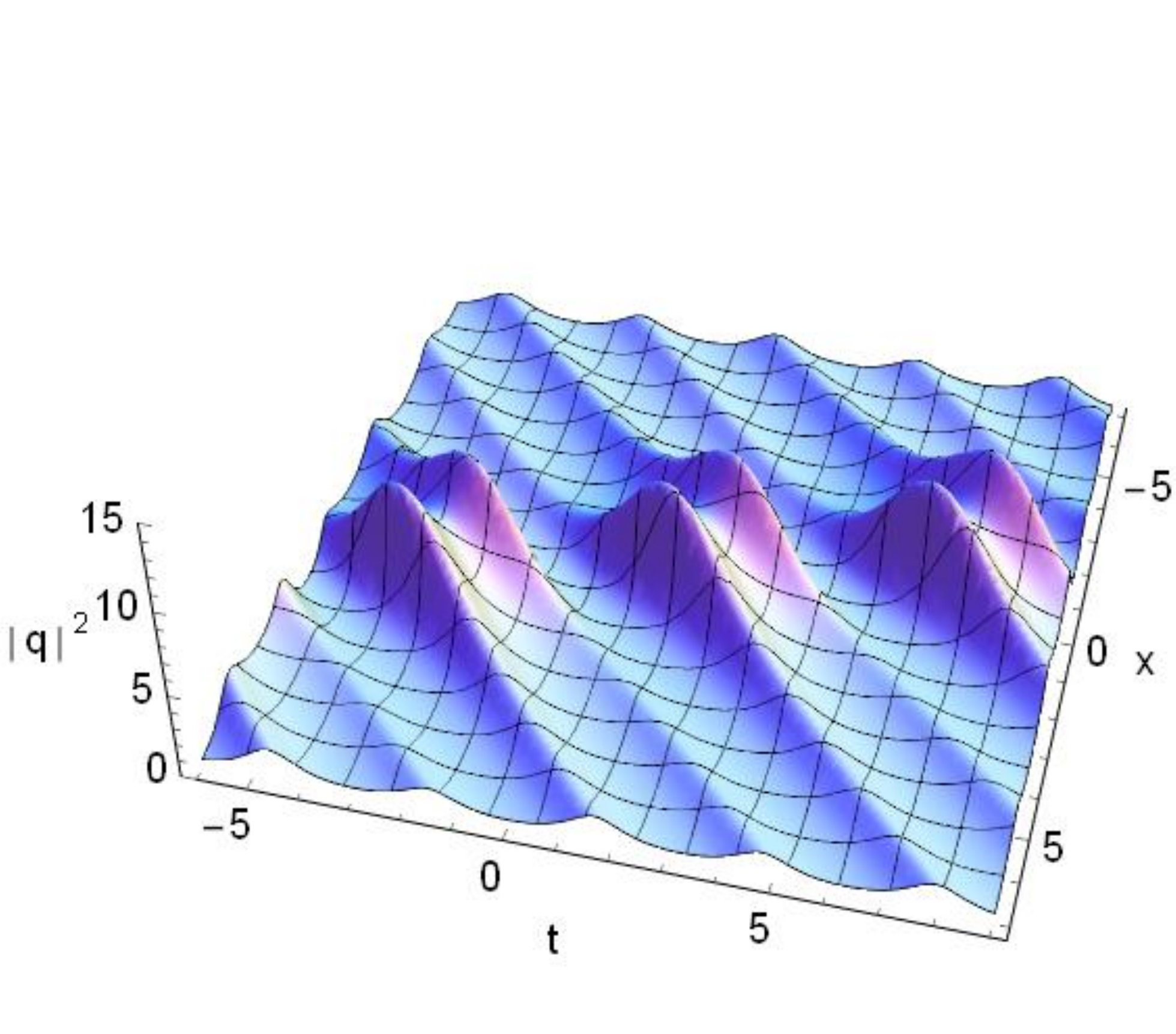}
  \end{minipage}
  }
  \hspace{1in}
  \subfigure[]{
  \begin{minipage}{4.5cm}
  \includegraphics[width=\textwidth]{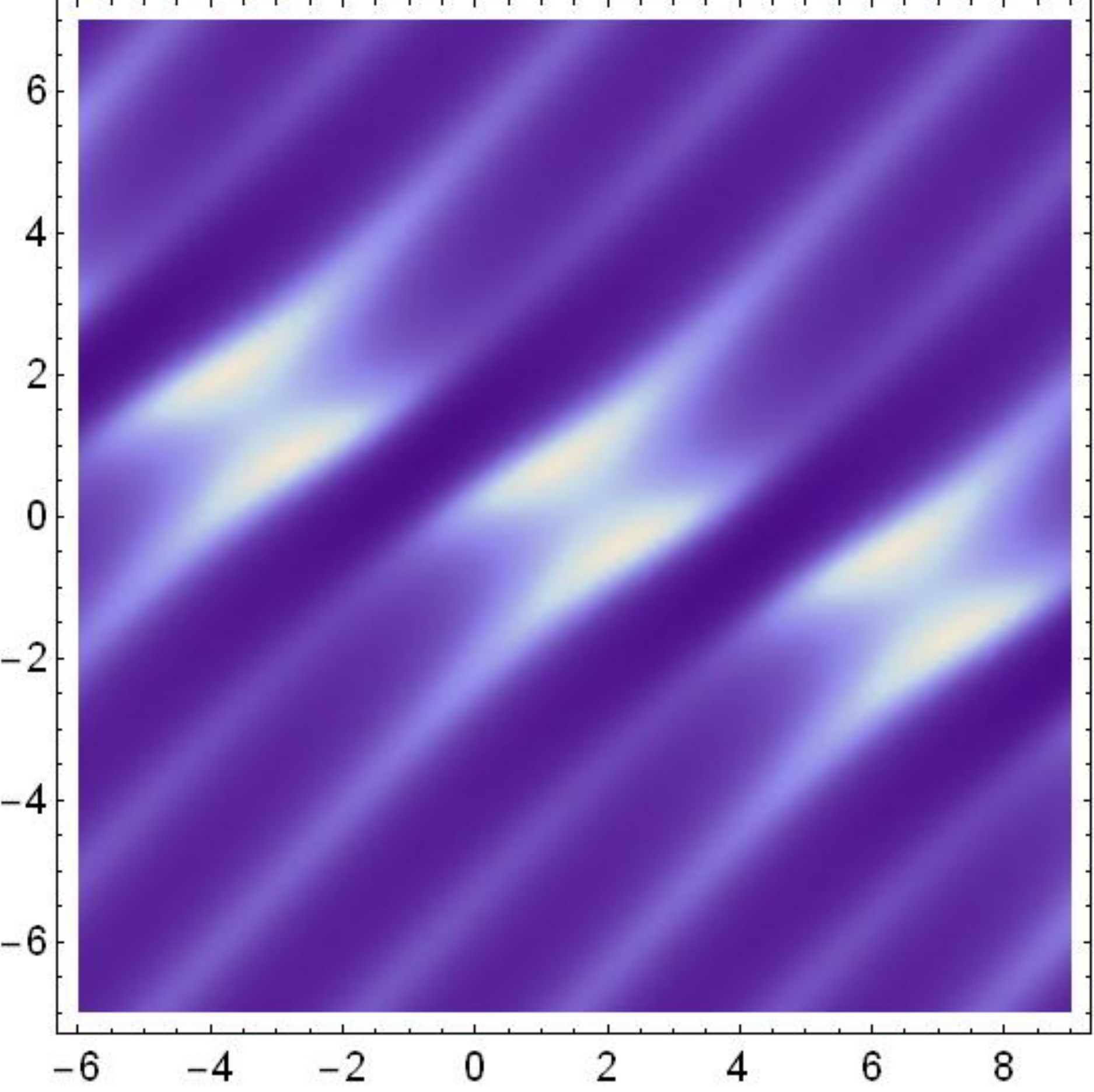}
  \end{minipage}
  }
  \caption{(a). Shape and motion of $|q|^2$ corresponding to \eqref{TA1} for equation \eqref{re-cmkdv-eq}, 
  in which $k_1=0.5, k_2=-0.5, h_1=i, x_0=1, t_0=1, c_1=2, c_2=1, d_1=2, d_2=1$.
  (b). Density plot of (a).}
  \label{fig1-2ab}
\end{figure}

\begin{figure}[ht]
      \centering
      \subfigure[]{
      \begin{minipage}{6.5cm}
      \includegraphics[width=\textwidth]{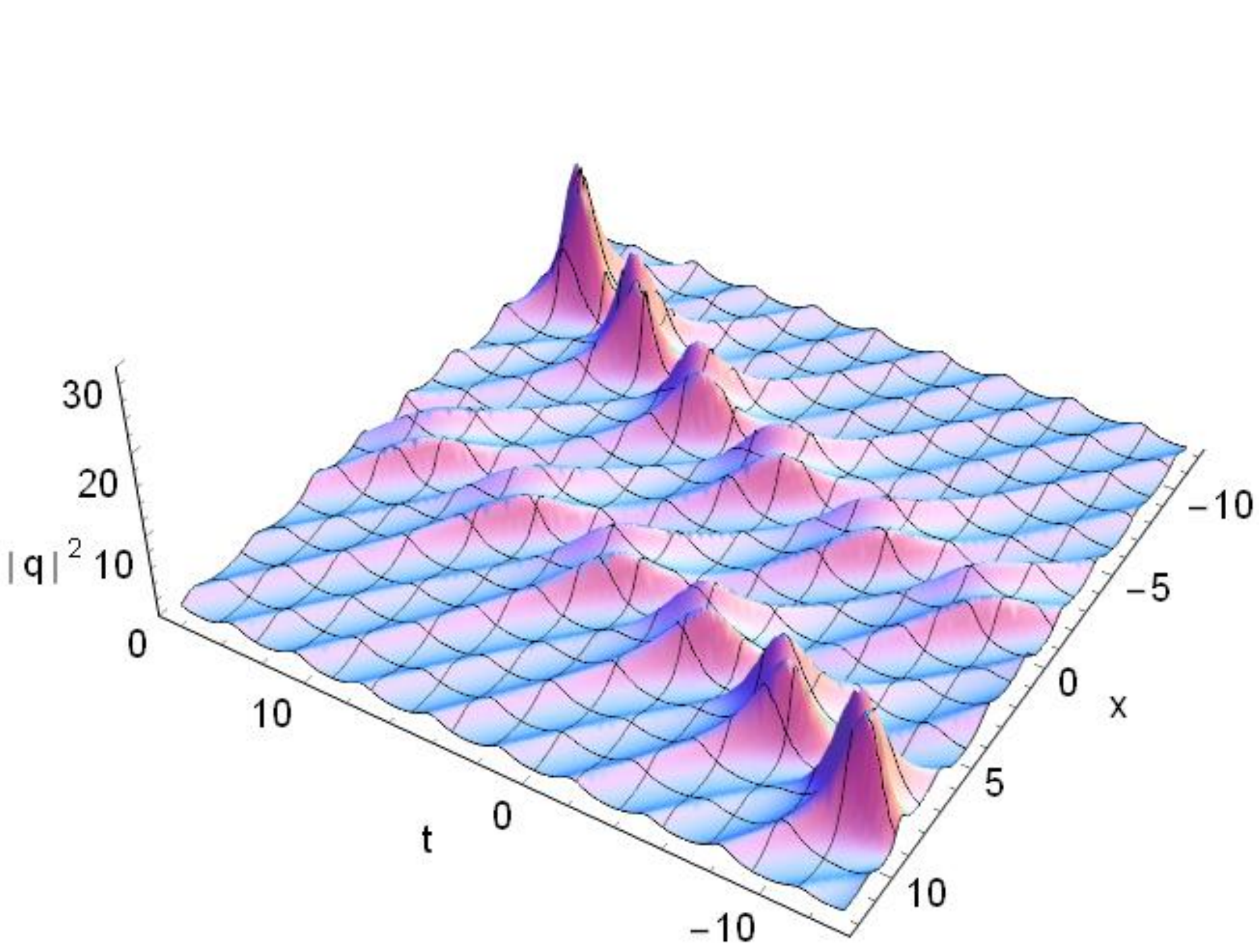}
      \end{minipage}
      }
      \hspace{1in}
      \subfigure[]{
      \begin{minipage}{4.5cm}
      \includegraphics[width=\textwidth]{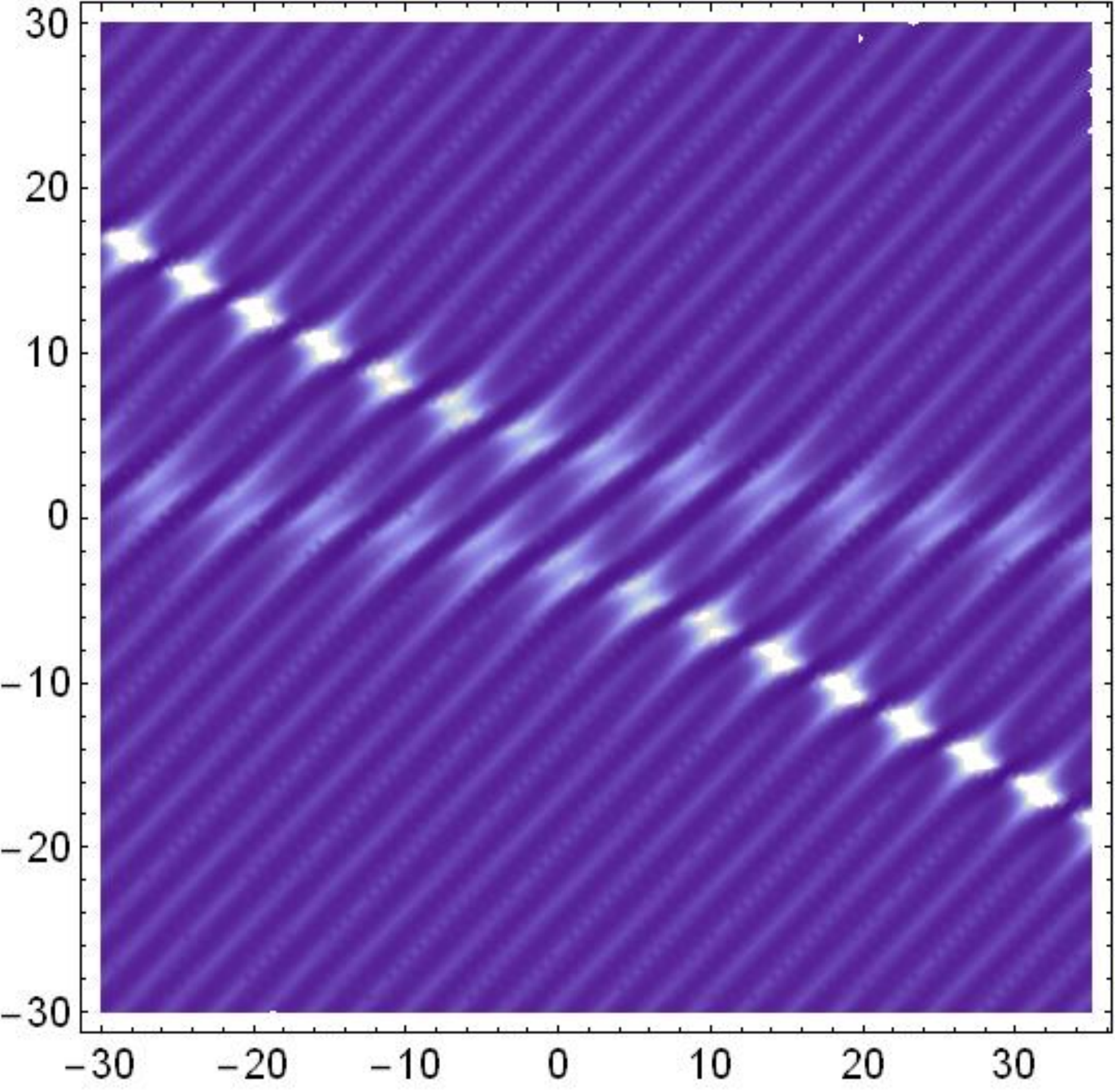}
      \end{minipage}
      }
      \caption{(a). Shape and motion of $|q|^2$ corresponding to \eqref{TA2} for equation \eqref{re-cmkdv-eq},
      in which $k_1=0.4, k_2=-0.4, k_3=-0.7, k_4=0.7, h_1=i, c^+_1= 4, c^+_2 = 1,
      c^+_3=2, c^+_4 =2, c^+_5 =2,c^+_6=1$.
               (b). Density plot of (a). }
       \label{fig1-2cd}
    \end{figure}

Based on the above analysis about one-soliton solutions, we consider mixed solutions.
For example, solitons with periodic backgrounds. Here we skip formulae of solutions and just illustrate them in
the following. Their explicit formulae can be easily written out from \eqref{mkdv-q-solu}.
Fig.\ref{fig1-2ab} corresponds to
  \begin{equation}\label{TA1}
    T=\left(\begin{array}{cccc}
    1& 0 & 0 & 0\\
    0 & -1 & 0 & 0 \\
    0 & 0 & 0 & 1\\
    0 & 0 & 1 & 0
    \end{array}
    \right),~~
    A=\left(\begin{array}{cccc}
    k_1 & 0 & 0 & 0\\
    0 & k_2 & 0 & 0 \\
    0 & 0 & h_1 & 0\\
    0 & 0 & 0 & h^*_1
    \end{array}
    \right),
    \end{equation}
and describes one soliton on a periodic background.
Fig.\ref{fig1-2cd} corresponds to
\begin{equation}\label{TA2}
  T=\left(\begin{array}{cccccc}
  1& 0 & 0 & 0&0&0\\
  0 & -1 & 0 & 0&0&0 \\
  0&0&1&0&0&0\\
  0&0&0&-1&0&0\\
  0 & 0 & 0&0&0 & 1\\
  0 & 0 &0&0& 1 &0
  \end{array}
  \right),~~
  A=\left(\begin{array}{cccccc}
  k_1 & 0 & 0 & 0 &0 &0\\
  0 & k_2 & 0 & 0 &0 &0\\
  0 & 0 & k_3 & 0&0 &0\\
  0 & 0 & 0 & k_4 &0 &0\\
  0 & 0 & 0 & 0 &h_1 &0\\
  0 & 0 & 0 &0 &0 &h^*_1
  \end{array}
  \right),
  \end{equation}
and describes interactions of two solitons on a periodic background.
In both cases $k_i \in\mathbb{R}$ and $h_1\in\mathbb{C}$.

\section{Solutions to the space-time shifted nonlocal semi-discrete NLS equation}\label{sec-3}

\subsection{Space-time shifted nonlocal semi-discrete NLS equation}\label{sec-3-1}

There exist space-time shifted nonlocal differential-difference (semi-discrete) integrable systems.
Consider the semi-discrete NLS equation (also known as the  AL equation)
\begin{equation}
i Q_{n,t}=Q_{n+1}+Q_{n-1}-2Q_n +|Q_n|^2(Q_{n+1}+Q_{n-1}),
\end{equation}
which is integrable and related to the
AL spectral problem \cite{AL-JMP-1975,Abl-PT-book-2004},
\begin{equation}\label{AL-sp}
\Theta_{n+1} =M_n  \Theta_n,~~
          M_n=\left(
           \begin{array}{cc}
             \lambda & Q_n \\
             R_n &  1/\lambda \\
           \end{array}
         \right),~~\Theta_n=\left(
           \begin{array}{cc}
             \theta_{1,n} \\
             \theta_{2,n} \\
           \end{array}
         \right),
\end{equation}
where $\lambda$ is a spectral parameter, $(Q_n, R_n)=(Q(n,t), R(n,t))$ are potential functions of
$(n,t)\in \mathbb{Z} \times \mathbb{R}$.

The unreduced coupled AL system reads \cite{AL-JMP-1975}
\begin{subequations}\label{nls-bef}
\begin{align}
& i Q_{n,t}=Q_{n+1}+Q_{n-1}-2Q_n -Q_nR_n(Q_{n+1}+Q_{n-1}),\\
& i R_{n,t}=-(R_{n+1}+R_{n-1}-2R_n) + Q_nR_n(R_{n+1}+R_{n-1}).
\end{align}
\end{subequations}
It admits the following space-time shifted nonlocal reductions,
\begin{subequations}\label{red-nls}
\begin{align}
& iQ_{n,t}=(Q_{n+1}+Q_{n-1}-2Q_n) -\delta Q_nQ_{n_0-n}^*(Q_{n+1}+Q_{n-1}),~~~ R_n=\delta Q_{n_0-n}^*,
\label{snls}\\
& iQ_{n,t}=(Q_{n+1}+Q_{n-1}-2Q_n)-\delta Q_nQ_{n_0-n}(t_0-t)(Q_{n+1}+Q_{n-1}),
~~~ R_n=\delta Q_{n_0-n}(t_0-t),
\label{tsnls}\\
& iQ_{n,t}=(Q_{n+1}+Q_{n-1}-2Q_n)-\delta Q_nQ_{n}(t_0-t)(Q_{n+1}+Q_{n-1}),~~~ R_n=\delta Q_n(t_0-t),
\label{tnls}
\end{align}
\end{subequations}
where $n_0, t_0$ are real parameters, $\delta=\pm1$ and
\[Q_{n_0-n}\doteq Q(n_0-n,t),~~ Q_{n_0-n}(t_0-t)\doteq Q(n_0-n,t_0-t)~~Q_n(t_0-t)\doteq Q(n,t_0-t).\]

\subsection{Solutions to the unreduced systems}\label{sec-3-2}

Let us list the double Casoratian solutions of the unreduced AL system obtained in \cite{DengLZ-AMC-2018}.

Through  transformation
\begin{eqnarray}\label{trans}
Q_n=\frac{g_n}{f_n},~~R_n=\frac{h_n}{f_n},
\end{eqnarray}
\eqref{nls-bef} is written into bilinear form
\begin{subequations}\label{nls-bi}
\begin{eqnarray}
  &&iD_{t}g_n\cdot f_n =g_{n+1}f_{n-1}+g_{n-1}f_{n+1}-2g_{n}f_{n}, \label{nls-bi-1}  \\
  &&iD_{t}f_n\cdot h_n =f_{n+1}h_{n-1}+f_{n-1}h_{n+1}-2f_{n}h_{n}, \label{nls-bi-2}  \\
  &&f_n^2-f_{n-1}f_{n+1}=g_nh_n,\label{nls-bi-3}
\end{eqnarray}
\end{subequations}
where $D$ is the  Hirota bilinear operator defined as before.
Introduce double Casoratian composed by $\Phi_n$ and $\Psi_n$ in terms of their double shifts,
\begin{align}
|\W {\Phi_n^{[N]}};\W {\Psi_n^{[M]}}|
%=\mathrm{Cas}^{[N+1,M+1]}(\Phi_n,\Psi_n)
=|\Phi_n,E^2\Phi_n,\cdots,E^{2N}\Phi_n; \Psi_n,E^2\Psi_n,\cdots,E^{2M}\Psi_n|,
 \end{align}
where
\[
\Phi_n=(\phi_{1,n},\phi_{2,n},\cdots,\phi_{N+M+2,n})^T,~~\Psi_n=(\psi_{1,n},\psi_{2,n},\cdots,\psi_{N+M+2,n})^T,
\]
$E$ is a shift operator defined by $E^j f_n=f_{n+j}$,
and
\begin{equation}\label{W-Phi}
\W {\Phi_n^{[N]}}=(\Phi_n,E^2\Phi_n,\cdots,E^{2N}\Phi_n).
\end{equation}

\begin{proposition}\label{P-5}\cite{DengLZ-AMC-2018}
The bilinear system \eqref{nls-bi} admits double Casoratian solutions
\begin{equation}
 f_n= |\W {\Phi_n^{[N]}};\W {\Psi_n^{[M]}}|, ~~ g_n= |\W {\Phi_n^{[N+1]}};\W {\Psi_n^{[M-1]}}|, ~~
  h_n= -|\W {\Phi_n^{[N-1]}};\W {\Psi_n^{[M+1]}}|,
 \label{fgh}
\end{equation}
where
\begin{eqnarray}\label{phi-psi-A}
\Phi_n=A^n\exp\Bigl(-\frac{i}{2}(A^2-2I +A^{-2})t\Bigr)C^+,~~
\Psi_n=A^{-n}\exp\Bigl(\frac{i}{2}(A^2-2I +A^{-2})t\Bigr)C^{-},
\end{eqnarray}
in which $A\in \mathbb{C}_{(N+M+2)\times(N+M+2)}$ is an invertible complex constant matrix,
$C^{\pm} \in \mathbb{C}_{N+M+2}$  and $I$ is the $(N+M+2)$-th order identity matrix.
\eqref{phi-psi-A} can be alternatively expressed as
\begin{eqnarray}\label{phi-B-nls}
\Phi_n=\exp\Bigl(nB -\frac{i}{2}(e^{2B}-2 I +e^{-2B})t\Bigr)C^+,~~
\Psi_n=\exp\Bigl(-nB +\frac{i}{2}(e^{2B}-2 I +e^{-2B})t\Bigr)C^{-},
\end{eqnarray}
where we have introduced
\begin{equation}\label{A-B}
A=e^B,
\end{equation}
with $B\in \mathbb{C}_{(N+M+2)\times (N+M+2)}$.
\end{proposition}

\subsection{Reductions of solutions to the space-time shifted nonlocal semi-discrete NLS equation}\label{sec-3-3}

\subsubsection{Double Casoratians to the reduced equations \eqref{red-nls}}\label{sec-3-3-1}

As in the continuous case, now let us explain how the reduction technique works in the Casoratian case.

For equation \eqref{snls} we take $M=N$ and replace $C^{\pm}$ by $A^{\mp N}C^{\pm}$ as they are arbitrary.
Then, in the following we consider double Casoratians
\begin{equation}\label{snls-fgh}
f_n=|A^{-N}\W {\Phi_n^{[N]}};A^{N}\W {\Psi_{n}^{[N]}}|,  ~~
g_n=|A^{-N}\W {\Phi_n^{[N+1]}};A^{N}\W {\Psi_{n}^{[N-1]}}|,  ~~
h_n= -|A^{-N}\W {\Phi_n^{[N-1]}};A^{N}\W {\Psi_{n}^{[N+1]}}|,
\end{equation}
which are still solution to \eqref{nls-bi}.
Introduce constraints
\begin{subequations}\label{snls-cons}
\begin{align}
& \Psi_n = T\Phi^*_{n_0-n}, \label{snls-cons-a}\\
&  C^-= TA^{*n_0}C^{+*}, \label{snls-cons-b}
\end{align}
\end{subequations}
where matrices $A$ and $T$ obey
\begin{subequations}\label{the-snls-AT}
\begin{align}
& AT-TA^*=0,\label{the-snls-AT-a} \\
& TT^*=-\delta|A^*|^2 I. \label{the-snls-AT-b}
\end{align}
\end{subequations}
Then, from \eqref{phi-psi-A} and making use of \eqref{snls-cons-b} and \eqref{the-snls-AT-a}, we find that
\begin{align*}
\Psi_n =&A^{-n}e^{\frac{i}{2}(A^2-2I +A^{-2})t}C^- \nonumber\\
=& (TA^*T^{-1})^ne^{\frac{i}{2}((TA^*T^{-1})^2-2I +(TA^*T^{-1})^{-2})t}TA^{*n_0}C^{+*} \nonumber\\
=& T A^{*(n_0-n)}e^{\frac{i}{2}(A^{*2}-2I +A^{*-2})t}C^{+*} \nonumber\\
=&T\Phi^*_{n_0-n},
\end{align*}
which means the constraint \eqref{snls-cons-a} coincides with \eqref{snls-cons-b} and \eqref{the-snls-AT-a}.
With \eqref{snls-cons-a} we write $f_n$ in \eqref{snls-fgh} as
\begin{equation}\label{ff}
f_n=|A^{-N}\W {\Phi_n^{[N]}};A^{N}T \W {\Phi_{n_0-n}^{*[N]}}|
=|A^{-N}\W {\Phi_n^{[N]}};TA^{*N}\W {\Phi_{n_0-n}^{*[N]}}|.
\end{equation}
Note that we have made use of $A^NT=TA^{*N}$ which is resulted from \eqref{the-snls-AT-a},
and we also specify that in light of \eqref{W-Phi},
\begin{equation*}
\W {\Phi_{n_0-n}^{*[N]}}=(\Phi_{n_0-n}^*,E^2\Phi_{n_0-n}^*,\cdots,E^{2N}\Phi_{n_0-n}^*)
=(\Phi_{n_0-n}^*,A^{*-2}\Phi_{n_0-n}^*,\cdots,A^{*-2N}\Phi_{n_0-n}^*).
\end{equation*}
Then, making use of  \eqref{the-snls-AT}, we can find that
\begin{align*}
f_n =&|A^{-N}\Phi_n,A^{-N+2}\Phi_n,\cdots,A^{N}\Phi_n;T {A^{*}}^{N}
\Phi^*_{n_0-n},T {A^{*}}^{N-2}\Phi^*_{n_0-n},\cdots,T {A^*}^{-N}\Phi_{n_0-n}^*|\\
=&|A^{-N}\W {\Phi_n^{[N]}};TA^{*N}\W {\Phi_{n_0-n}^{*[N]}}|\\
=&(-\delta|A^*|^{-2})^{N+1}|T|| T^*A^{-N}\W {\Phi_n^{[N]}};A^{*N}\W {\Phi_{n_0-n}^{*[N]}}|   \\
=&(\delta|A^*|^{-2})^{N+1}|T|| A^{*N}\W {\Phi_{n_0-n}^{*[N]}};T^*A^{-N}\W {\Phi_n^{[N]}}|   \\
=&(\delta|A^*|^{-2})^{N+1}|T||A^N\!\Phi_{n_0-n},A^{N-2}\Phi_{n_0-n},\cdots,A^{-N}\!\Phi_{n_0-n};
                                 T\! {A^*}^{-N}\!\Phi^*_{n},T\!{A^*}^{-N+2}\Phi^*_{n},
                                 \cdots,T\!{A^{*}}^{N}\Phi_{n}^*|^* \\
=&(\delta|A^*|^{-2})^{N+1}|T||A^N\Phi_{n_0-n},A^{N-2}\Phi_{n_0-n},\cdots,A^{-N}\Phi_{n_0-n};
                                 {A}^{-N}T\Phi^*_{n},{A}^{-N+2}T\Phi^*_{n},\cdots,{A}^{N}T\Phi_{n}^*|^* \\
=&(\delta|A^*|^{-2})^{N+1}|T||A^{-N}\Phi_{n_0-n},A^{-N+2}\Phi_{n_0-n},\cdots,A^N\Phi_{n_0-n};
                                 {A}^{N}T\Phi_{n}^*,{A}^{N-2}T\Phi^*_{n},\cdots,{A}^{-N}T\Phi^*_{n}|^* \\
=&(\delta|A^*|^{-2})^{N+1}|T|f_{n_0-n}^*.
\end{align*}
Similarly, we have
\[h_n=\delta^{N}|A^*|^{-2(N+1)}|T|g_{n_0-n}^*,\]
which yields
\begin{equation}\label{snls-fgh'}
R_n=\frac{h_n}{f_n}=\frac{\delta^{N}|A^*|^{-2(N+1)}|T|g_{n_0-n}^*}
{(\delta|A^*|^{-2})^{N+1}|T|f_{n_0-n}^*}=\delta\frac{g_{n_0-n}^*}{f_{n_0-n}^*}=\delta Q^*_{n_0-n},
\end{equation}
Thus, when we take  \eqref{snls-cons-a} together  with \eqref{snls-cons-b} and \eqref{the-snls-AT-a},
from \eqref{snls-fgh'} and $Q_n=g_n/f_n$ we get solution to the reduced equation \eqref{snls}.
In a similar way we can implement reductions and obtain solutions to the reduced equations \eqref{tsnls} and \eqref{tnls}.
Let us summarize these results below.

\begin{theorem}\label{T-3}
The nonlocal semi-discrete NLS equation \eqref{snls} and \eqref{tsnls} allow double Casoratian solutions
\begin{equation}\label{the-snls-Q}
 Q_n=\frac{g_{n}}{f_{n}},
\end{equation}
with
\begin{equation}
 f_n=|A^{-N}_{}\W {\Phi_n^{[N]}};A^{N}_{}\W {\Psi_{n}^{[N]}}|,~~
 g_{n}=|A^{-N}\W {\Phi_n^{[N+1]}};A^{N}\W {\Psi_{n}^{[N-1]}}|,
\end{equation}
and $\Phi_n$   given  in \eqref{phi-psi-A} or equivalently in \eqref{phi-B-nls},
where for equation \eqref{snls},
\begin{align}\label{snls-cons-a'}
 \Psi_n = T\Phi^*_{n_0-n},
\end{align}
and $A$ and $T$ obey  the relation
\begin{align}
 AT-TA^*=0, ~~
 TT^*=-\delta|A^*|^2 I, \label{the-snls-AT'}
\end{align}
or, in terms of $B$,
\begin{equation}\label{the-snls-BT}
BT-TB^*=0, ~~TT^*=-\delta|e^{B^*}|^2 I,
\end{equation}
and where for equation \eqref{tsnls},
\begin{equation}\label{tsnls-cons}
 \Psi_n = T\Phi_{n_0-n}(-t),
\end{equation}
and $A$ and $T$ obey
\begin{equation}\label{the-tsnls-AT}
AT-TA=0,~~T^2=-\delta|A|^2 I,
\end{equation}
or equivalently
\begin{equation}\label{the-tsnls-BT}
BT-TB=0, ~~T^2=-\delta|e^B|^2 I.
\end{equation}
Equation \eqref{tnls} admits solution \eqref{the-snls-Q} with
\begin{equation}\label{the-tnls-Q}
f_n=|\W {\Phi_n^{[N]}};\W {\Psi_n^{[N]}}|,~~
 g_n=|\W {\Phi_n^{[N+1]}};\W {\Psi_n^{[N-1]}}|,
\end{equation}
where $\Phi_n$ is given as in \eqref{phi-psi-A} or in \eqref{phi-B-nls},
\begin{equation}\label{tnls-cons}
 \Psi_n = T\Phi_n(t_0-t),
\end{equation}
$A$ and $T$ obey the relation
\begin{equation}\label{the-tnls-AT}
A^{-1}T-TA=0, ~~ T^2=\delta I,
\end{equation}
or equivalently,
\begin{equation}\label{the-tnls-BT}
BT+TB=0, ~~T^2= \delta I.
\end{equation}
\end{theorem}

Note that for equation \eqref{tsnls} and equation \eqref{tnls}, $C^-$ takes as
\[C^-=TA^{n_0}\exp\Bigl(-\frac{i}{2}(A^2-2I +A^{-2})t_0\Bigr)C^+\]
and
\[C^-=Te\Bigl(-\frac{i}{2}(A^2-2I +A^{-2})t_0\Bigr)C^+,\]
respectively.

\subsubsection{Solutions and examples}\label{sec-3-3-2}

Let us give solutions to the matrix equations \eqref{the-snls-BT}, \eqref{the-tsnls-BT} and \eqref{the-tnls-BT}.

\begin{proposition}\label{P-6}
(1). For equation \eqref{the-snls-BT}, when $|e^{B}|\in \mathbb{R}$ we can assume
\begin{equation}\label{BT}
B= \left(
      \begin{array}{cc}
        K_1 & \mathbf{0} \\
        \mathbf{0}& K_4
      \end{array}
    \right),~~
    T=\alpha \left(
      \begin{array}{cc}
         T_1 & T_2 \\
         T_3 &  T_4
      \end{array}
    \right),
\end{equation}
where $T_j, K_j\in \mathbb{C}_{(N+1)\times (N+1)}$, $\alpha=|e^{B}|\in \mathbb{R}$.
Then, $K_j$ and $T_j$ can be given as in Proposition \ref{P-3}.\\
(2). For equation \eqref{the-tsnls-BT}, we assume \eqref{BT} where $\alpha=|e^{B}|$.
Then, $K_j$ and $T_j$ can be given as in Table \ref{Tab-1} where $\sigma=-1$.\\
(3). For equation \eqref{the-tnls-BT}, we assume \eqref{BT} where $\alpha=1$.
Then, $K_j$ and $T_j$ can be given as in Table \ref{Tab-1} where $\sigma=1$.
\end{proposition}

Explicit forms of $\Phi_n$ can be given accordingly.
Define
\begin{equation}\label{theta}
\theta(k)=kn-2it \sinh^2k.
\end{equation}
Then, for example, when
\[K_1=\mathrm{Diag}[k_i]_{i=1}^{N+1},~~ K_4=\mathrm{Diag}[h_i]_{i=1}^{N+1},\]
we have
\[\Phi_n=\Bigl(c_{1}e^{\theta(k_{1})},c_{2}e^{\theta(k_{2})},\cdots,c_{N+1}e^{\theta(k_{N+1})},
  d_{1}e^{\theta(h_1)},d_{2}e^{\theta(h_2)},\cdots,d_{N+1}e^{\theta(h_{N+1})}\Bigr)^{T},\]
and when
\[K_1=\mathbf{J}_{N+1}[k],~~ K_4=\mathbf{J}_{N+1}[h],\]
we have
\[  \Phi=\Bigl(ce^{\theta(k)},\frac{\partial_{k}}{1!}(ce^{\theta(k)}), \cdots,
  \frac{\partial_{k}^{N}}{N!}(ce^{\theta(k)}),
  de^{\theta (h)},\frac{\partial_{h}}{1!}(de^{\theta (h)}),\cdots,
  \frac{\partial_{h}^{N}}{N!}(de^{\theta (h)})\Bigr)^{T},
\]
where $c_j, d_j, c, d \in \mathbb{C}$.
Explicit forms of $\Phi_n$ of other cases can easily be given as in Sec.\ref{sec-2-4-1}
according to different canonical forms of $B$.
Note that  $B$ (or $A$) and its any similar matrix lead to same solutions to $Q_n$ and $R_n$ through \eqref{trans}.

In the following, as examples we list out one-soliton solutions of the three nonlocal equations
\eqref{snls}, \eqref{tsnls} and \eqref{tnls}.

For equation \eqref{snls}, it has two one-soliton solutions.
One is for $\delta=\pm 1$,
\begin{equation}\label{sol-3-1}
Q_n=\frac{(e^{2k_1}-e^{2k_1^*})c_1c_2}
{e^{k_1+k_1^*}(|c_2|^2 e^{-2\theta(k_1)+n_0k_1}+\delta |c_1|^2 e^{-2\theta(k_1^*)+n_0k_1^*}},~~
k_1, c_j \in \mathbb{C},
\end{equation}
and the other is for only $\delta=-1$,
\begin{equation}\label{sol-3-2}
Q_n=\frac{(e^{2k_1}-e^{2h_1})c_1c_2}
{e^{k_1+h_1}(c_1^*c_2 e^{-2\theta(k_1)+n_0k_1}+c_1c_2^*e^{-2\theta(h_1)+n_0h_1})},~~k_1,h_1 
\in \mathbb{R},~
c_j\in \mathbb{C},
\end{equation}
where $\theta(k)$ is given as in \eqref{theta}. For equation \eqref{tsnls}, its one-soliton is
\begin{equation}
Q_n=\frac1{\sqrt{-\delta}}\frac{e^{2k_1}-e^{2h_1}}
{e^{k_1+h_1}(e^{-2\theta(k_1)+n_0k_1-2i t_0\sinh^2k_1}+e^{-2\theta(h_1)+n_0h_1-2i t_0\sinh^2h_1})},
~~k_1, h_1 \in \mathbb{C}.
\end{equation}
For equation \eqref{tnls}, its one-soliton is
\begin{equation}\label{sol-4-1}
Q_n=\frac1{\sqrt{-\delta}}\frac{(e^{2k_1}-e^{-2k_1})e^{-4it\sinh^2 k_1}c_1c_2}
{(c_1^2e^{2k_1n}+c_2^2e^{-2k_1n})e^{-2it_0\sinh^2 k_1}},~~ k_1, c_j \in \mathbb{C}.
\end{equation}

\begin{figure}[ht]
  \centering
  \subfigure[]{
  \begin{minipage}{6.0cm}
  \includegraphics[width=\textwidth]{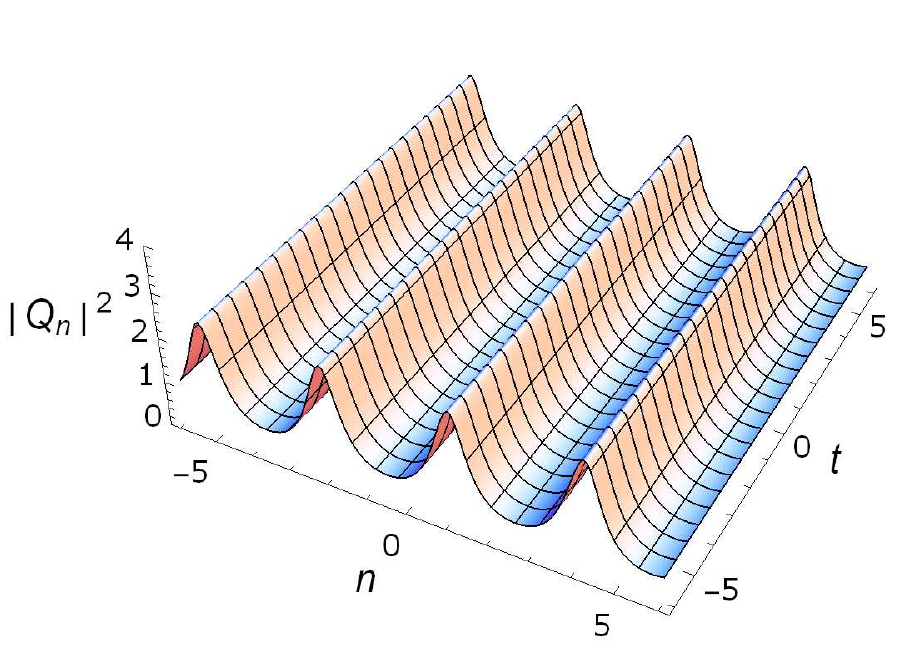}
  \end{minipage}
  }
  \hspace{-2mm}
  \subfigure[]{
  \begin{minipage}{6.0cm}
  \includegraphics[width=\textwidth]{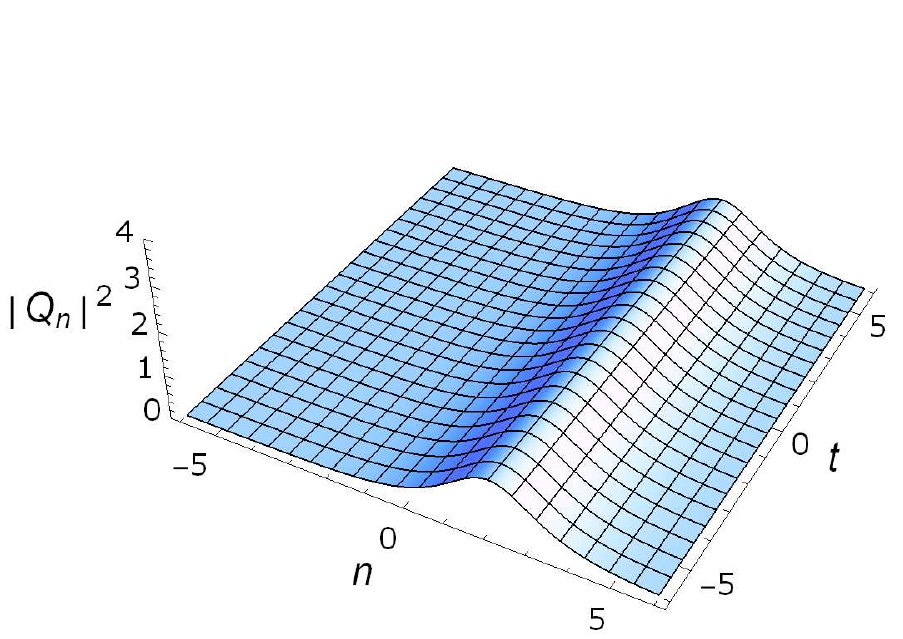}
  \end{minipage}
  }
  \caption{(a). Shape and motion of the solution \eqref{sol-snls-1} for equation \eqref{snls}, 
  in which $k_1=0.5i, n_0=2, c_1=0.5,c_2=0.3$.
  (b). Shape and motion of the solution \eqref{sol-snls-2} for equation \eqref{snls}, 
  in which $k_1=0.5, h_1=-0.5,n_0=4, c_1=1,c_2=1$.}

  \label{fig-4}
\end{figure}

Finally,  let us briefly look at dynamics of these solutions and identify the roles of $n_0$ and $t_0$.
As for solution \eqref{sol-3-1} to equation \eqref{snls}, the wave package gives rise to
\begin{equation}\label{sol-snls-1}
|Q_n|^2=\frac{4|c_1c_2|^2e^{4na-2n_0a}\sin^22b}{(e^{16st}|c_1|^4+e^{-16st}|c_2|^4)
+2\delta |c_1c_2|^2\cos (4nb-2n_0b)}.
\end{equation}
Here and after we take $k_1=a+ib$ with $a,b \in \mathbb{R}$, $s=4\sinh 2a \cos 2b$.
It provides a nonsingular periodic solution
under the case $a=0, |c_1|\neq |c_2|$,
as shown in Fig.\ref{fig-4}(a),
for the rest, it has periodic singularities  at
\begin{align*}
& n=\frac{m\pi}{2b}+\frac{n_0}{2},~~ t=\frac{1}{8s}\ln|\frac{c_2}{c_1}|, ~~~\delta=-1;\\
& n=\frac{(2m+1)\pi}{4b}+\frac{n_0}{2},~~ t=\frac{1}{8s}\ln|\frac{c_2}{c_1}|, ~~~\delta=1,
\end{align*}
where $m\in \mathbb{Z}$.
$n_0$   brings changes of amplitude and the location of singularities.
As for solution \eqref{sol-3-2} to equation \eqref{snls} with $\delta=-1$,
the wave package reads
\begin{equation}
|Q_n|^2=\frac{(e^{2k_1}-e^{2h_1})^2e^{-2(k_1+h_1)}}
{e^{-4k_1n+2n_0k_1}+e^{-4h_1n+2n_0h_1}+2e^{-2(k_1+h_1)n+n_0(k_1+h_1)}
\sin [\gamma-4(\sinh^2k_1-\sinh^2h_1)t]},
\end{equation}
where $\gamma=\arctan \frac{\mathrm{Re}[(c_1^*c_2)^2]}{\mathrm{Im}[(c_1^*c_2)^2]}$.
It is  singular except  the special case $h_1=-k_1$ and $\sin \gamma \neq -1$, i.e.,
\begin{equation}\label{sol-snls-2}
  |Q_n|^2=\frac{2\sinh ^2(2k_1)}{\cosh (4k_1n-2n_0k_1)+\sin \gamma},
  \end{equation}
which gives rise to a stationary soliton 
with constant amplitude $\frac{2\sinh ^2(2k_1)}{1+\sin \gamma}$,
as shown in Fig.\ref{fig-4}(b).

As for the solution \eqref{sol-4-1} to the equation \eqref{tnls}, we have
\begin{equation}\label{sol-Q-4}
|Q_n|^2=\frac{4[(\sinh 2a \cos 2b)^2+(\cosh 2a \sin 2b)^2]e^{16st-4st_0}}
{|c_2|^{-2}e^{4na}+|c_1|^{-2}e^{-4na}+2\sin (\gamma-4nb)},
\end{equation}
here $s$ and $\gamma$ are defined as before.
From equation \eqref{sol-Q-4}, we can easily see that $t_0$
affects only the phase of wave.
With regard to  dynamics, when $|c_1c_2|<1$, \eqref{sol-Q-4}
is a nonsingular periodic wave but its amplitude  exponentially changes with time.
There are two special cases of this solution.
One is for $k_1$ being real, i.e., $k_1=a$, solution \eqref{sol-Q-4} reads
\begin{equation}\label{sol-Q-4-a}
|Q_n|^2=\frac{4\sinh^2 2a}{|c_2|^{-2}e^{4na}+|c_1|^{-2}e^{-4na}+2\sin \gamma},
\end{equation}
which is a stationary soliton;
the other is $k_1$ being pure imaginary, i.e., $k_1=ib$, solution \eqref{sol-Q-4} reduces to
\begin{equation}\label{sol-Q-4-b}
|Q_n|^2=\frac{4\sin^2 2b}{|c_2|^{-2}+|c_1|^{-2}+2\sin (\gamma-4nb)},
\end{equation}
which is a stationary periodic wave with period $\frac{\pi}{2b}$ in space.

\section{Conclusions}\label{sec-4}

In this paper, by means of a reduction technique based on bilinearization and double Wronskians/Casoratians,
we derived explicit multi-soliton solutions for the space-time shifted nonlocal NLS and mKdV hierarchies
and the semi-discrete space-time shifted nonlocal NLS equation.
In this approach we made use of double Wronskian/Caosratian solutions of the unreduced  systems,
to convert nonlocal reductions to the constraints to the elementary vectors $\varphi$ and $\psi$ in  double Wronskians
(and $\Phi_n$ and $\Psi_n$ in double Caosratians),
which require the eigenvalue matrix $A$ (and $B$ in discrete case)
satisfies some constrained matrix equations, e.g. \eqref{nls1-at-t-eq}, \eqref{nls2-at-t-equation} 
and \eqref{nls3-at-t-equation}
(and  \eqref{the-snls-BT}, \eqref{the-tsnls-BT} and \eqref{the-tnls-BT} for discrete case).
As we have seen that, compared with the nonlocal equations without space-time shifts,
the  distributions of eigenvalues do not change with  space-time shifts but the space-time shifts
do bring new constraints to the phase terms in solutions.
This observation will be helpful for the investigation of the nonlocal space-time shifted integrable equations
using other approaches, such as the inverse scattering transform and Darboux transformation.

Finally, we remark that two-soliton solutions of the space-time shifted nonlocal NLS equation and mKdV equation
were obtained in a recent paper \cite{GP-arxiv-2021}.
However, in the present paper our reduction technique enables us to derive solutions to a hierarchy of equations,
present distributions of eigenvalues and obtain explicit formulae of
multi-soliton solutions and multiple-pole solutions.

\subsection*{Acknowledgments}
This project is supported by the NSF of China (Nos. 11631007 and 11875040).

\end{document}